\documentclass[lettersize,journal]{IEEEtran}
\usepackage{amsmath,amsfonts}
\usepackage{algorithmic}
\usepackage{algorithm}
\usepackage{array}
\usepackage[caption=false,font=normalsize,labelfont=sf,textfont=sf]{subfig}

\usepackage{cite}
\usepackage{xcolor}

\usepackage[caption=false]{subfig}
\usepackage[font=small]{caption}
\usepackage{sansmath}
\DeclareCaptionFont{sansmath}{\sansmath}
\usepackage{booktabs}
\usepackage{xspace}
\usepackage{markdown}

\newcommand{\vdd}{$V_{\rm DD}$\xspace}

\newcommand{\vds}{$V_{\rm DS}$\xspace}

\newcommand{\vbg}{$V_{\rm BG}$\xspace}

\newcommand{\vt}{$V_t$\xspace}

\newcommand{\idvg}{$I_{\rm DS}-V_{\rm GS}$\xspace}
\newcommand{\TCryo}{$T_{\rm Cryo}$\xspace}

\newcommand{\tapas}[1]{{\color{black} #1}}

\usepackage[colorlinks]{hyperref}
\hypersetup{%
  colorlinks=true,
  linkcolor={blue},
  citecolor={red},
  urlcolor={blue},
  }

\begin{document}

\bstctlcite{IEEEexample:BSTcontrol} 

\title{A Methodology for Process Design Kit Re-Centering Using TCAD and Experimental Data for Cryogenic Temperatures}

\author{Tapas Dutta, Fikru Adamu-Lema, Djamel Bensouiah, and Asen Asenov
\thanks{T. Dutta, F. Adamu-Lema, and A. Asenov are with Semiwise Ltd., Glasgow, and also with the Device Modelling Group, University of Glasgow, UK. D. Bensouiah is with Semiwise Ltd., Glasgow.
(e-mail: tapas.dutta@glasgow.ac.uk)}}

\markboth{}%
{Dutta \MakeLowercase{\textit{et al.}}: A Methodology for Process Design Kit Re-Centering Using TCAD and Experimental Data for Cryogenic Temperatures}


\maketitle
\begin{abstract}
In this work, we describe and demonstrate a novel Technology Computer Aided Design (TCAD) driven methodology to re-center room-temperature Process Design Kits (PDKs) for cryogenic operation using a limited set of experimental measurements. Unlike previous approaches that relied on direct fitting of sparse measurements, our technique accounts for process-induced deviations by calibrating TCAD models to both room-temperature and cryogenic data. Compact models for all process corners are extracted from TCAD-generated target characteristics, enabling accurate cryogenic modeling without dedicated foundry support.  This scalable, technology-independent method provides a practical path for cryogenic circuit design.

\end{abstract}

\section{Introduction}


Cryogenic electronics has attracted growing interest in both academia and industry in recent times. This emerging focus spans applications in high-performance computing (HPC), quantum technologies, and data center efficiency. For example, operating conventional CMOS at deep cryogenic temperatures (e.g., 77 K or below) has been proposed as a way to significantly reduce leakage power and thermal noise in large-scale data centers and HPC systems\cite{SaligramChip24,MorozSISPAD21}. Likewise, in quantum computing, integrating control and readout circuits at cryogenic temperatures (4–77 K) is critical to eliminate the wiring and latency bottlenecks that arise when interfacing cold qubits with room-temperature electronics \cite{XueNature21,patraJSSC18, GalyJEDS18}. These trends have driven the development of novel cryo-CMOS design methodologies, supported by recent advances in cryogenic circuit design and compact model development \cite{CharbonIEDM16,AkturkBCICTS24}.

These trends underscore the need for reliable cryogenic PDKs to enable complex circuit design under cryogenic conditions. An important hurdle is that the foundry PDKs are designed for room temperature operation and at the time of executing this work, and to the best of our knowledge, no foundry-provided PDKs are available for design at cryo temperature. Measurements at cryogenic temperatures can be used to re-center\cite{WangSISPAD16} the room temperature PDK to cryogenic temperatures. However, this is a complicated process due to discrepancies between the characteristics of the Typical/Typical (TT) transistors from the foundry PDK and the transistors measured on the silicon wafers. The foundry provides no guarantee  to customers that the fabricated  devices will have the same characteristics as the TT transistors in the PDK. Instead, the foundry only guarantees that the transistor characteristics on the fabricated wafers will be in-between the characteristics of the fast/fast (FF) and slow/slow (SS) corner transistor characteristics in the PDK. 

While several methods have been proposed to address cryogenic CMOS modeling, they usually are limited to direct extraction of compact models using data from a few cryogenic temperature measurements of one or in some cases a few devices. Some such recent efforts are: cryogenic modeling of 28nm bulk CMOS tuning two parameters of the PSP model\cite{JungICEIC24}, BSIM-CMG model extraction for FinFETs\cite{SinghLAEDC24} etc. However, the simplistic approach of fitting limited measurements and extracting a compact model cannot produce a cryogenic PDK as the measured die can be anywhere between FF, SS corners, or even outside. While for room temperature (RT) models, measurements across many die/wafers/lots are performed and PDK is extracted, but for cryogenic CMOS this is not as easy and is also much more expensive (an example of cryogenic PDK from scratch has been reported in\cite{AkturkBCICTS24}, but that is for an old 130nm CMOS technology). Further, this simplistic approach has been usually targeted at a specific device architecture or technology node, which is useful, but has limited applicability for real circuit design, in contrast to the generic approach that we have developed. In summary, this highlights that existing one-off modeling approaches, although valuable for proof-of-concept, fall short of providing the comprehensive device libraries that engineers need for robust cryogenic circuit design.

This paper outlines a procedure to accurately re-center room temperature foundry PDKs for supporting circuit design at cryogenic temperatures, using a combination of a minimum set of experimental cryogenic measurements of CMOS transistors on test chips and TCAD simulations. The methodology delivers PDK quality models for TT, SS, FF, SF and FS transistors and include statistical variability (mismatch). A flowchart summarizing the entire methodology is shown in Fig. 1 (corners skipped for simplicity). The next sections describe the various steps involved in re-centering procedure. Note that we have demonstrated the methodology for standard 22nm FDSOI technology at 77K and 4K temperatures, isothermal TCAD with no back gate bias applied, and excluding PEX/BEOL parasitic extraction. Further, while the compact-model used in this implementation is BSIM-IMG, alternative FD-SOI models such as L-UTSOI \cite{PoirouxTED15} could also be used within the same flow. Here, we focus on the enablement of digital/analog DC circuit operation; RF modeling demonstration is outside the present scope. All figures in this work show nMOS devices; the methodology is identical for pMOS.

\begin{figure*}[!htbp]
\centering
\includegraphics[width=0.9\textwidth]{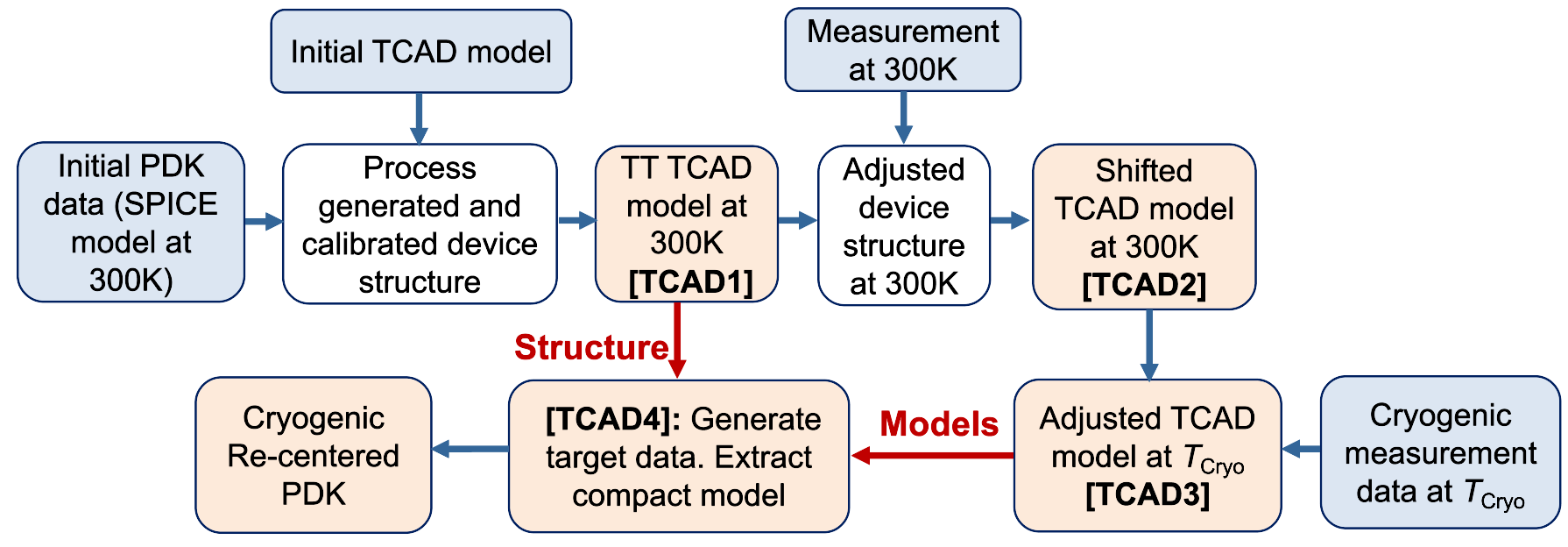} 
\caption{Overview of the cryogenic PDK re-centering methodology. The flowchart outlines the sequential steps starting from TCAD structure creation and calibration using room-temperature PDK data, followed by adjustment using measured data from fabricated devices, and culminating in the generation of compact models for cryogenic temperatures. The approach supports TT, SS, and FF corner models and can be extended to include statistical variability. (Corner-specific branches are omitted here for clarity.).\vspace{-3mm}}
\label{fig:recenteringFlow}
\end{figure*}

\begin{figure}[!htbp]
\centering
\captionsetup[subfloat]
{captionskip=-2em,margin=1.2em,justification=raggedright,singlelinecheck=false,font=small}
\subfloat[]{\includegraphics[width=0.33\textwidth,trim={0cm 0cm 0.0cm 0cm},clip]{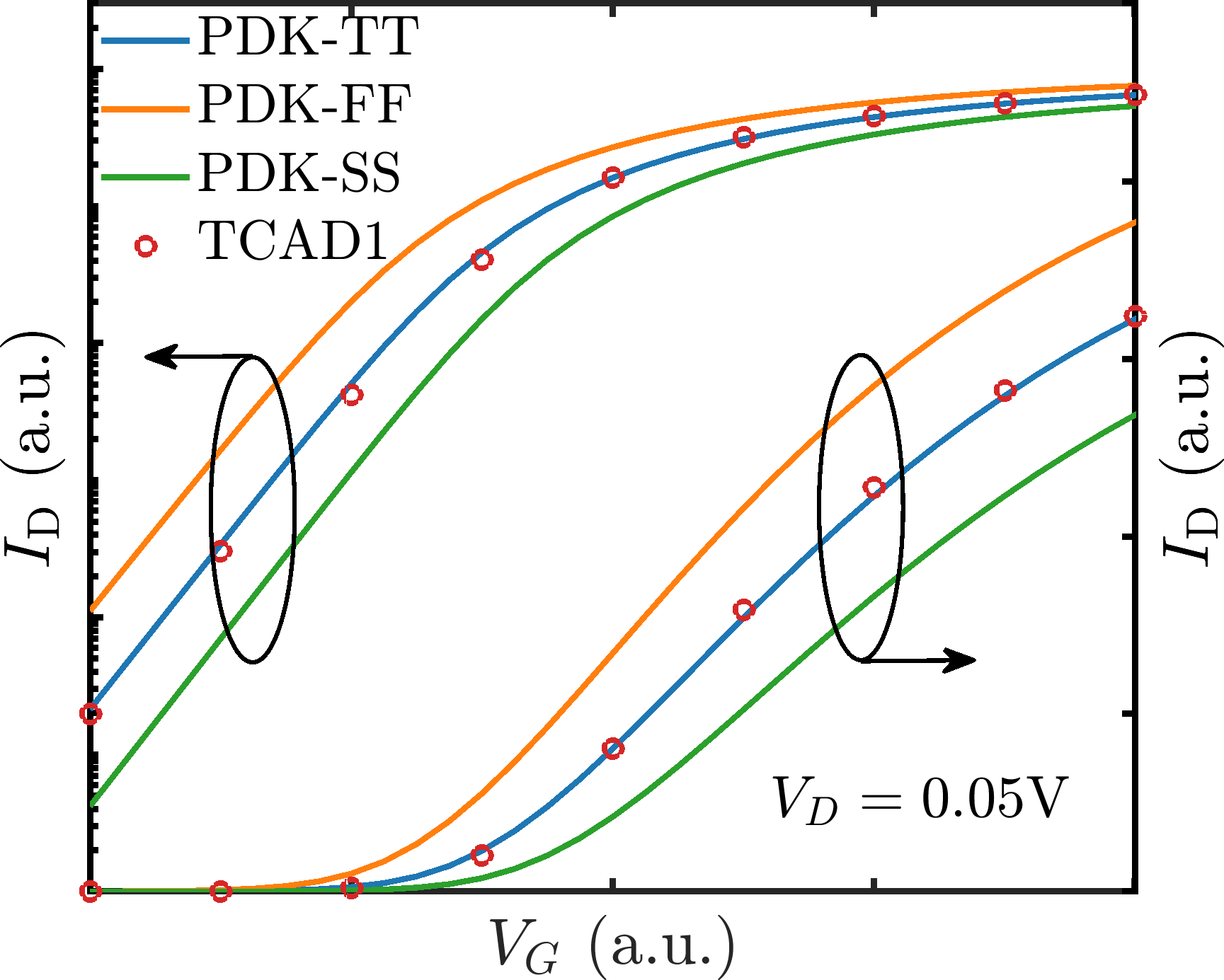}}

\subfloat[]{\includegraphics[width=0.33\textwidth,trim={0cm 0cm 0.0cm 0cm},clip]{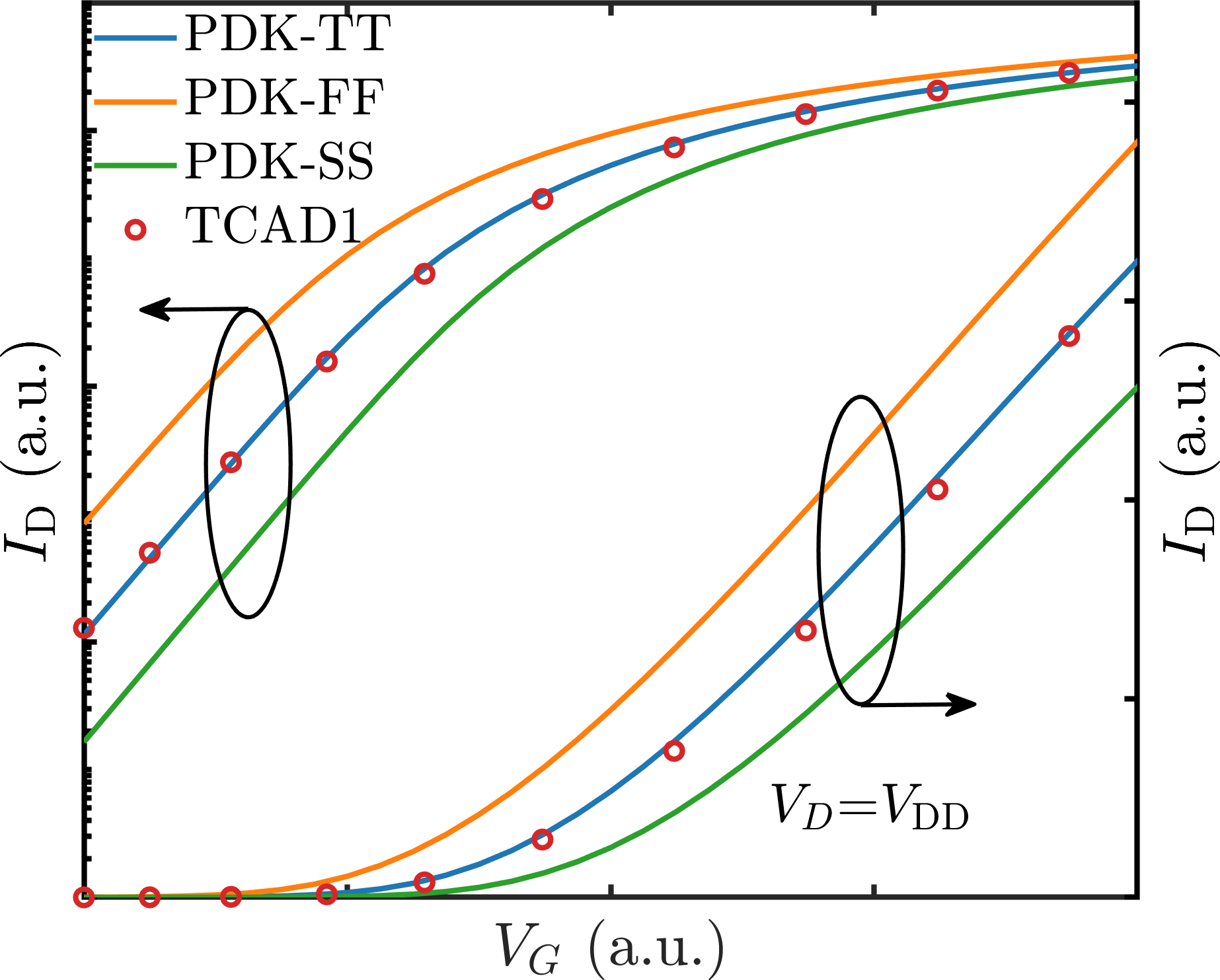}}
\caption{Comparison of calibrated TCAD derived characteristics of the TT transistor with foundry-provided PDK data at room temperature (T = 300 K): (a) linear (\vds=0.05V) (b) saturation (\vds=\vdd). Also shown are the slow-slow (SS) and fast-fast (FF) corner transistors from the room-temperature PDK for reference. Back gate bias, \vbg=0V. The units used are arbitrary (a.u.).}
\label{fig:calib}
\end{figure}

\section{TCAD Deck Calibration for TT Transistor}
The initial step in the PDK Re-Centering process is the creation of a TCAD deck, starting with the generation of a TCAD device structure of the typical-typical (TT) transistor, considering the room temperature PDK data (SPICE model generated). The structure is created using process simulations\cite{SentaurusProcess}), and device simulations are performed under the drift-diffusion formalism\cite{SentaurusManual22} with Fermi-Dirac carrier statistics.. This step includes device simulations along with the adjustments of the doping profiles and the transistor structure to match the electrostatic behavior of the TT transistors, which includes the threshold voltage (\vt), the Subthreshold Slope (SS), the Drain Induced Barrier Lowering (DIBL) at different bias conditions and their dependence on the transistor dimensions. This is followed by carrier mobility calibration aiming to represent accurately the transistor performance at low and high drain biases. The mobility models (Philips unified mobility model\cite{KlassenSSE92},
Lombardi model\cite{Lombardi}, and
Hansch \cite{HanschSSE89} model)
are selected to capture the temperature dependence as accurately as possible. An example of such a TCAD simulated characteristics obtained for a TT $n$-channel transistor calibrated to TT transistor data from the PDK is illustrated in Fig. 2. For comparison, the characteristics of the slow-slow (SS) and the fast-fast (FF) transistors as obtained from the room temperature PDK are also plotted in the figure. For this particular demo, we have used 22nm FDSOI devices\cite{CarterIEDM16}, but the methodology presented is generic and not dependent on a particular device architecture, technology node, or compact model. It should be noted that TCAD simulations can face convergence difficulties at deep-cryogenic temperatures. At such conditions, the combination of extremely low intrinsic carrier concentrations, partial dopant ionization, and sharp Fermi–Dirac distributions makes the coupled Poisson-continuity equations numerically stiff, which can hinder convergence in Newton-Raphson solvers. To tackle this, enhanced numerical settings (e.g. extended precision) and modified solver strategies (e.g. use of slow transient approach for the quasi-stationary voltage sweeps) are used in the TCAD simulator.

\section{TT Transistor TCAD Model Adjustment}
Next, we compare the room temperature characteristics of the TT transistor from TCAD (Fig. 2) and the corresponding measured data from a fabricated silicon wafer as illustrated in Fig. 3 (a). As expected, the measured transistor data is shifted with respect to the corresponding TT transistor. The TT transistor represents the average transistor characteristics across the wafer, across the lots and from lot to lot. The actual transistor characteristics on each wafer are different from the average transistor characteristics due to uncontrollable variations in the fabrication conditions.The main process parameters that cause the process variation are the dose and energy of different implantation steps, the gate oxide thickness, the annealing temperatures, and the transistor dimensions. These sources of variation are well established in CMOS device and process literature\cite{Sze,TaurNing,AsenovTED03}. Typically, up to 5\%-10\% variations in the above process parameters are expected during the fabrication process. The different technology parameters have different impact on threshold voltage, electrostatic integrity and drive current and should be carefully tested and combined to give the desirable shift in transistor behavior from the TT to the measured transistor. \tapas{The result from the simulation of the process calibrated structure (in order to match the room temperature measurement) is illustrated in Fig. 3 (b).} The new TCAD deck represents a Shifted TT (STT) transistor, which will be used at the next stage to perform calibration of the mobility, density gradient (for quantum correction), and other cryogenic specific models at cryogenic temperatures. This calibrated STT device provides a solid foundation for the subsequent cryogenic adjustment, as it captures process-dependent offsets present in the room-temperature PDK.

\begin{figure}[!htbp]
\centering
\captionsetup[subfloat]{captionskip=-2.6em,margin=1.6em,justification=raggedright,singlelinecheck=false,font=normal}
\subfloat[]{\includegraphics[width=0.33\textwidth,trim={0cm 0cm 0.0cm 0cm},clip]{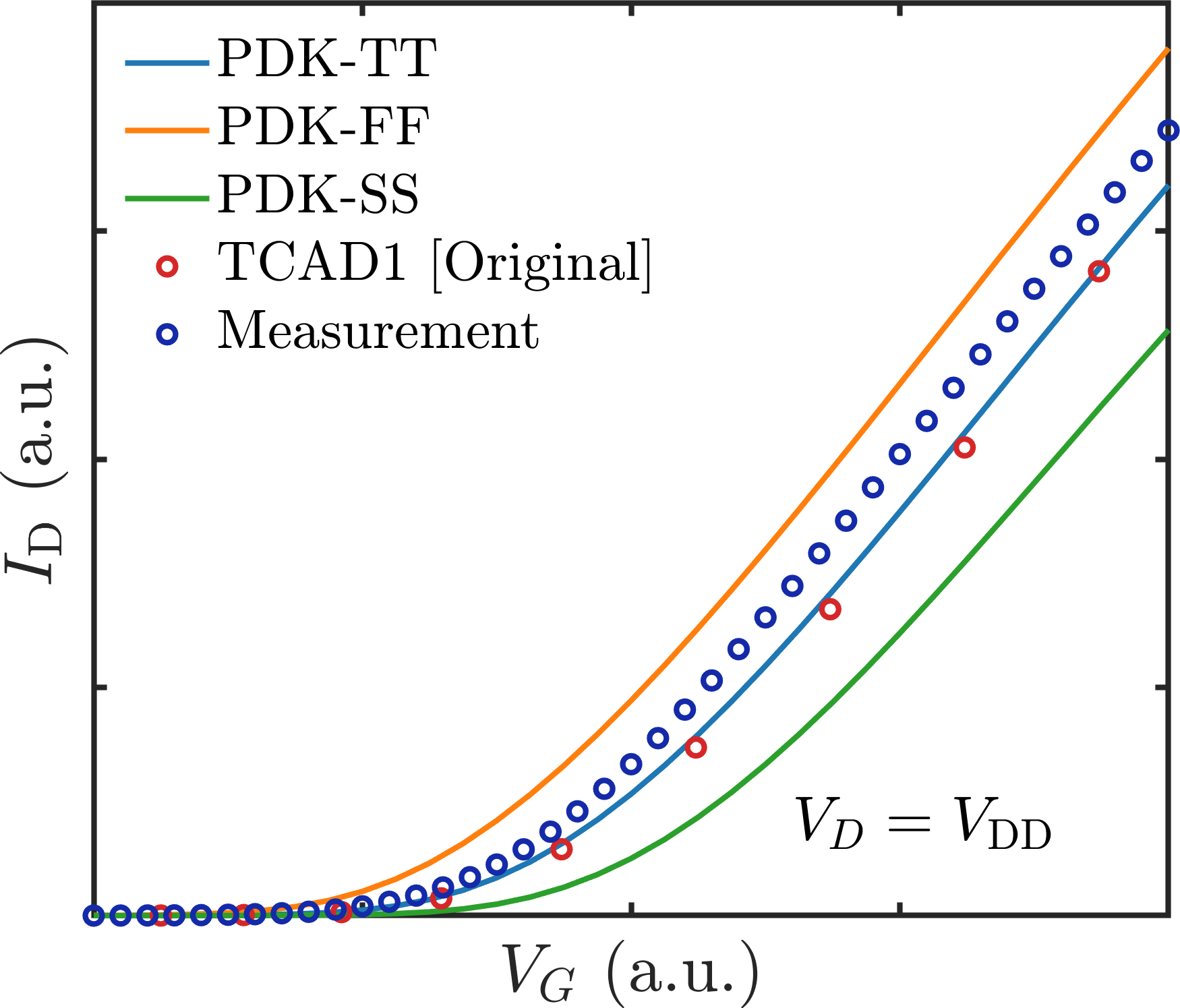}}

\subfloat[]{\includegraphics[width=0.33\textwidth,trim={0cm 0cm 0.0cm 0cm},clip]{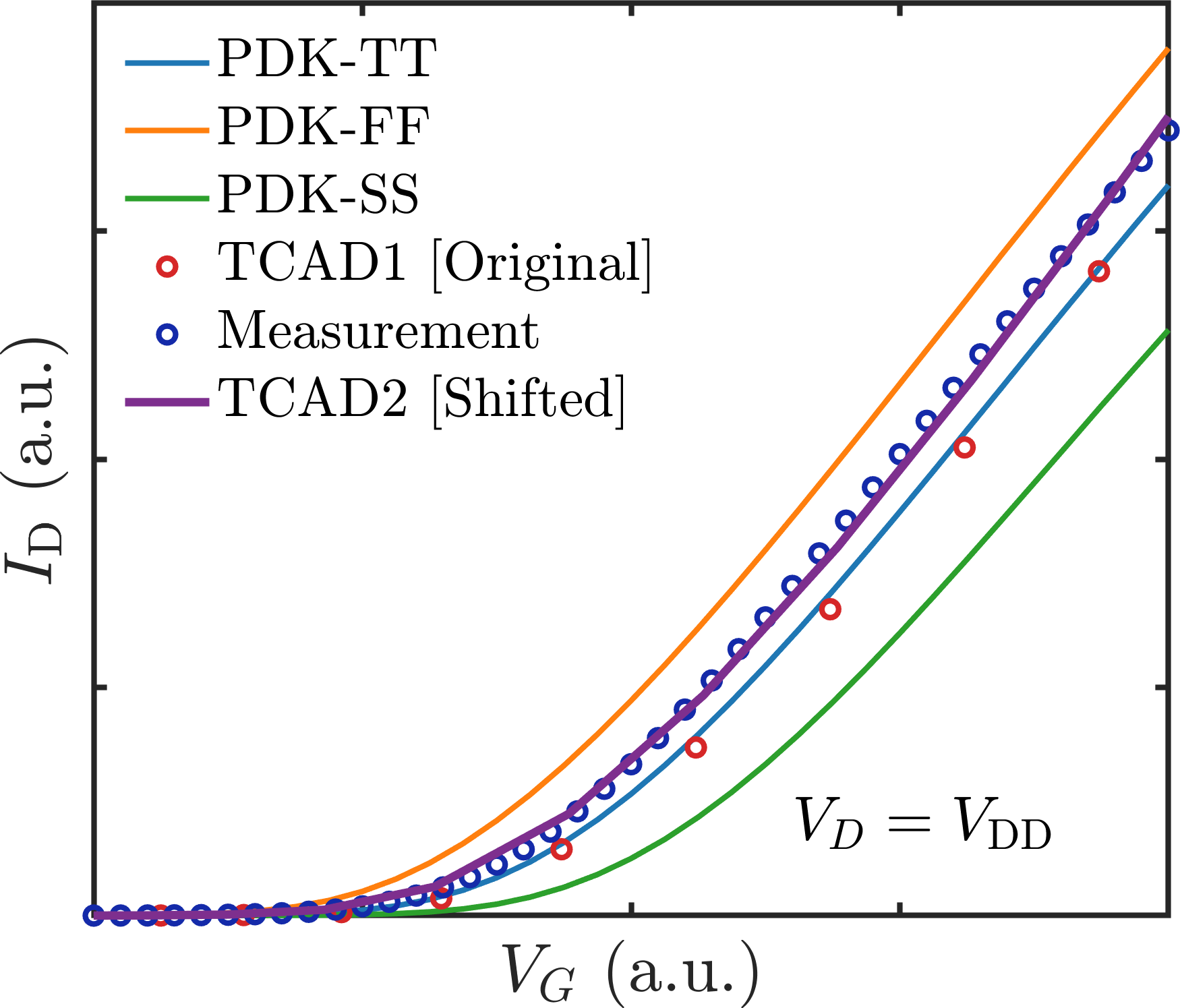}}

\subfloat[]{\includegraphics[width=0.36\textwidth,trim={0cm 0cm 0.0cm 0cm},clip]{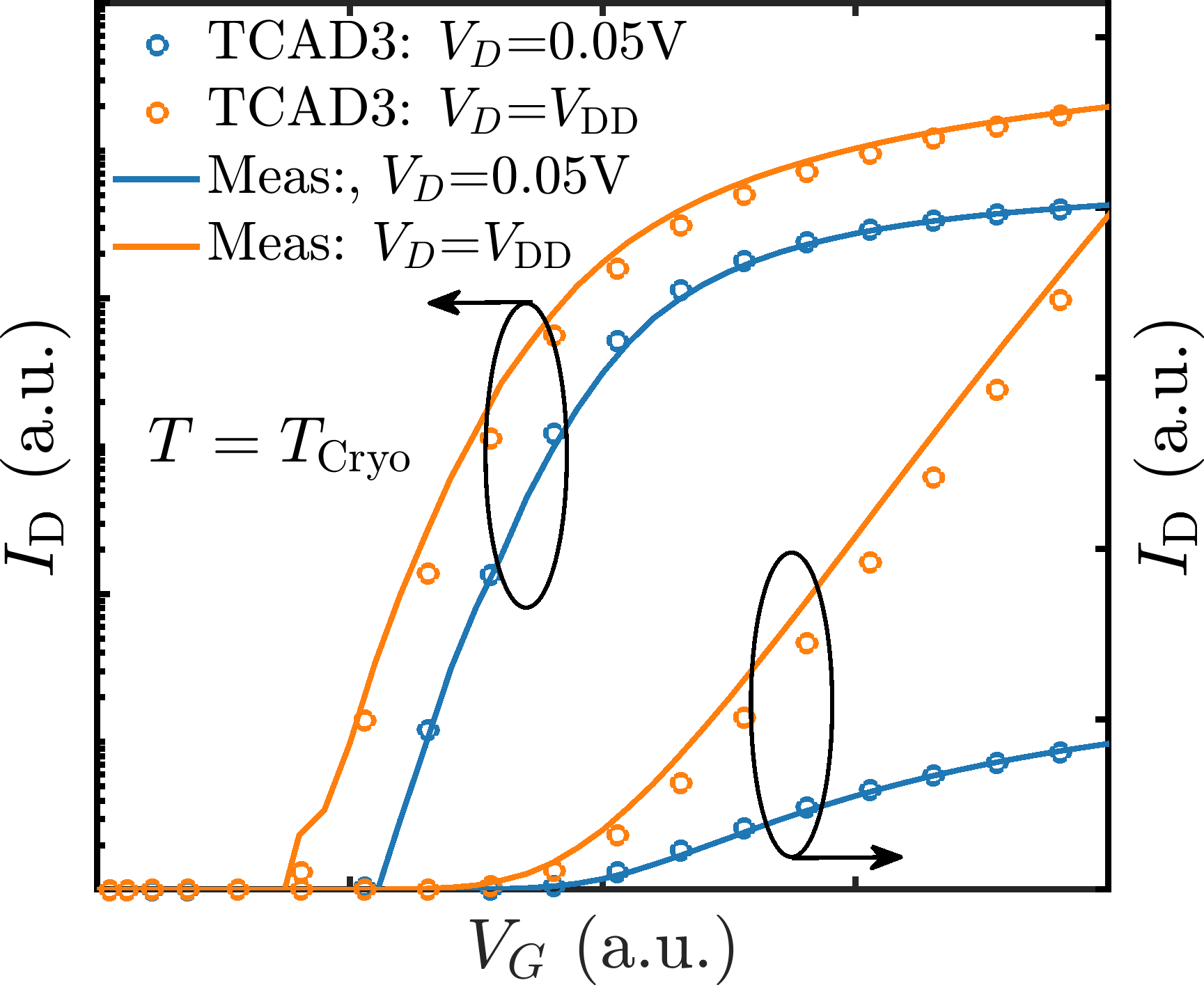}}

\caption{(a) Comparison of \idvg characteristics of the TT transistor in Fig. 2 and the corresponding measurement data at T=300K, highlighting the mismatch (b) Comparison and match between the measurement data on a fabricated silicon wafer and the calibrated shifted TCAD model. T=300K. (c) Comparison between the measurement data and the calibrated shifted TCAD model both at cryogenic temperature ($T$=\TCryo) at low and high drain biases. \vbg=0V.}
\label{fig:shifted}
\end{figure}

\section{Adjusted Transistor at Cryogenic Temperature}
At this stage, the STT TCAD transistor simulations are calibrated to the cryogenic transistor measurements. The process includes two steps – electrostatic calibration and mobility calibration. The electrostatic calibration aims to reproduce \vt, SS, and DIBL of the STT at the cryogenic temperature (\TCryo). In this paper, \TCryo = 77K, although we have applied the methodology at 4K as well. Considering the impact of the band tail states\cite{KanePR63} was critical for calibrating the subthreshold characteristics. We have used the Gaussian formulation of the band tail density of states, as implemented in \cite{SentaurusManual22}. 
By modeling band-tail state occupation at cryogenic temperatures, the TCAD simulation can replicate the measured subthreshold swing and off-state leakage current, which would otherwise be overly optimistic if one assumed an ideal abrupt band edge. 

The mobility calibration follows the calibration procedure at room temperature aiming to reproduce the current voltage characteristics above and below threshold. At room temperature, phonon scattering dominates mobility, but as temperature drops, lattice vibrations freeze out and carriers experience much less phonon scattering. As a result, the low-field mobility tends to increase at cryo \cite{BeckersTED18, PazVLSIT20}. However, other scattering mechanisms, such as Coulomb scattering off ionized impurities and surface roughness scattering in the channel begin to play a comparatively larger role\cite{TakagiTED94, MamunTED22}. The results from the calibration of the STT transistor at \TCryo are illustrated in Fig. 3 (c). This cryogenic mobility calibration ensured that both the subthreshold region and strong inversion region of the device’s I–V behavior were accurately captured by the TCAD model, providing confidence that the subsequent compact model extraction would reflect true device performance.

\section{Target Data Generation at Cryogenic Temperature and Model Extraction}
After the calibration of the STT transistor, the TCAD structure of the ‘original’ TT transistor is used along with the calibrated band tail, incomplete ionization and mobility models at cryogenic temperature from the previous step to generate the target characteristics for the compact model\cite{BsimOnline} extraction of the TT transistor at cryogenic temperature. The model cards for the compact model are extracted to yield the cryogenic PDK using standard compact model extraction procedures. The generated target characteristics at cryogenic temperature and the extracted compact model (HSPICE \cite{HSPICEManual22} results) of the TT transistor are illustrated in Fig. 4.

\begin{figure}[!htbp]
\centering
\includegraphics[width=0.35\textwidth]{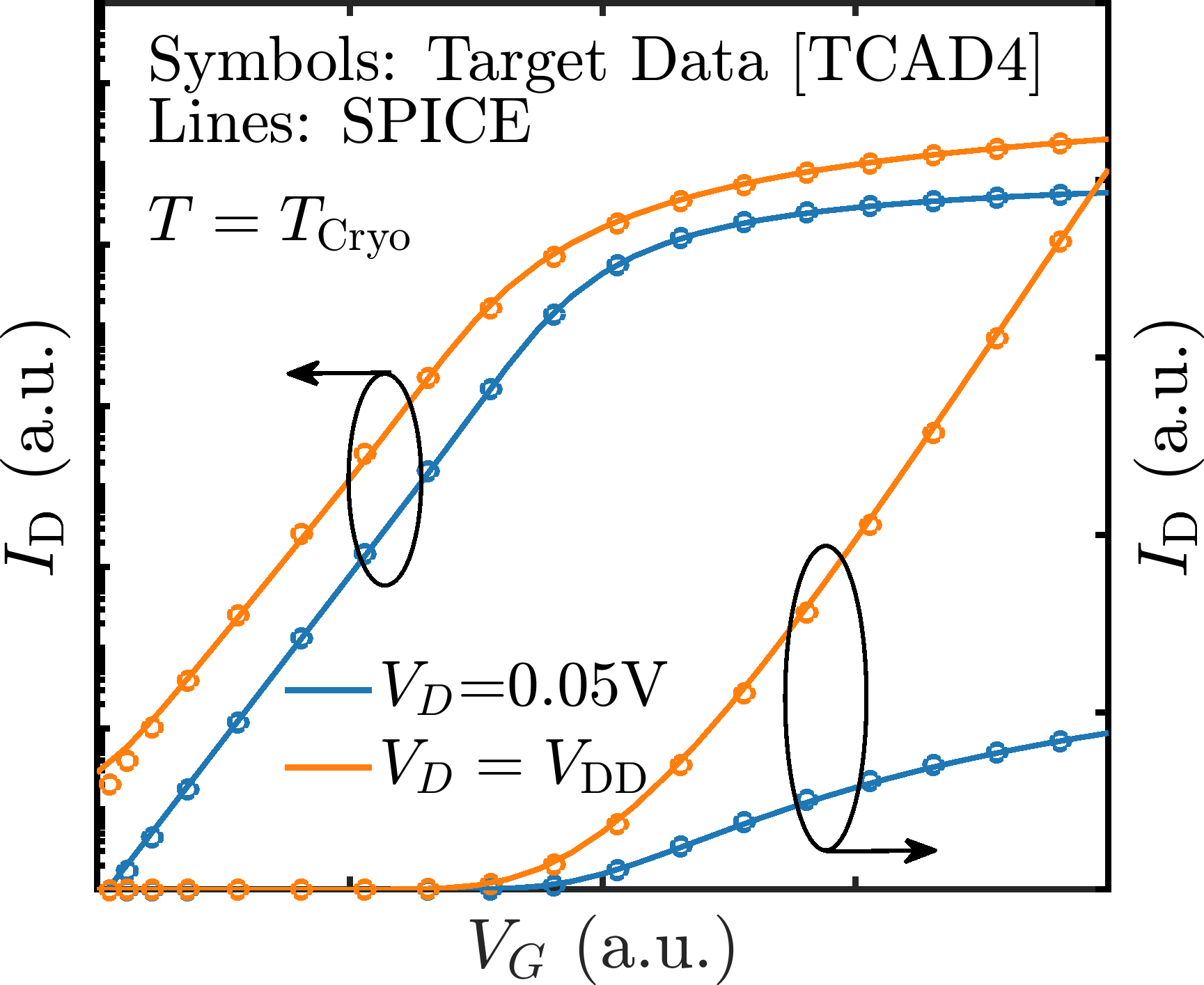} 
\caption{Comparison between the target \idvg data obtained from TCAD simulation of the original TT transistor at cryogenic temperature and the extracted compact model at low and high drain biases. \vbg=0V.}
\label{fig:TargetIdVg}
\end{figure}

The same approach is applied to the FF and SS corners. We have also developed a Monte-Carlo (MC) approach to consider variability in the cryogenic PDK, which will be described in a future publication.


\section{Discussion}
While the proposed methodology provides a systematic path to re-centering room-temperature PDKs for cryogenic operation, there are certain limitations, in particular the issue of accuracy of TCAD models at cryogenic temperatures. Commercial TCAD tools are primarily optimized for room-temperature device simulation. At deep cryogenic temperatures, convergence can become challenging, and the accuracy of built-in physical models can be limited. Although we have used state of the art TCAD models that target specificities of cryogenic operation (e.g. band tail model etc.) but residual discrepancies can’t possibly be completely avoided. Further, our compact model parameter extraction methodology does not attempt to re-extract the full set of compact model parameters. Instead, we target a limited subset which are enough to fit the target data. This strategy reduces complexity but may neglect secondary effects which will be considered in future extensions. 

\section{Conclusions}
In this paper we have described a new approach for PDK re-centering based on TCAD and experimental data at cryogenic temperatures suitable for circuit design applications in data centers and quantum computers. A similar procedure can be used to derive target data for cryogenic compact model extraction for SS and FF corners. Further work is in progress to extend this methodology for deriving statistical SPICE models, which are essential for verifying circuit performance at cryogenic temperatures. 

While the methodology presented here is generic and can be ported to future semiconductor nodes and device architectures, we have demonstrated the approach for DC PDK re-centering at 77 K (and 4 K, reported elsewhere) on 22 nm FDSOI, using \vbg=0 V, isothermal device simulations, and no PEX/BEOL parasitics. Once a minimal cryo measurement set is available, the end-to-end calibration and extraction typically requires days to weeks of engineering effort, depending on the number of corners/geometries targeted. Extending the flow to include self-heating, back-gate bias dependence, AC/capacitances, and RF can be taken up as future work.

Our methodology is applicable to other CMOS flavors like FinFETs and beyond, essentially any process where a baseline PDK exists and TCAD tools can model the low-temperature physics. This means that as industry moves to ever smaller geometries (or even novel device structures like nanowires and nanosheets), one can generate cryogenic PDKs for those platforms without starting from scratch. In the long term, we envision that readily available cryogenic PDKs will accelerate innovation in cryogenic computing. By lowering the barrier to designing complex CMOS systems at cryogenic temperatures, this work can help accelerate new advancements in energy-efficient supercomputing and scalable quantum processors, bridging the gap between room-temperature design capabilities and the needs of emerging cryogenic applications.

\section*{Acknowledgement}
We are grateful to GlobalFoundries for providing the 22FDX PDK and allowing us to customize it for cryogenic temperature operation. The device measurements were performed by Incize SRL, Belgium. This work was supported partially by Innovate UK funded project ``Development of Cryo-CMOS to enable the next generation of scalable quantum computers” under the grant number of 10006017 and was also partially supported by Semiwise Ltd, UK.

\bibliographystyle{IEEEtranDOI}
\bibliography{biblio.bib}

\end{document}